\documentclass[conference]{IEEEtran}

\usepackage{cite}
\usepackage{amsmath,amssymb,amsfonts}
\usepackage{algorithmic}
\usepackage{graphicx}
\usepackage{textcomp}
\usepackage{xcolor}
\usepackage[utf8]{inputenc} 
\usepackage[T1]{fontenc}    
\usepackage{url}            
\usepackage{booktabs}       
\usepackage{tabularx}
\usepackage{hyperref}
\usepackage{array}
\usepackage{subcaption}
\usepackage{multirow}
\usepackage{nicefrac}       
\usepackage{microtype}      
\usepackage{csquotes}
\usepackage{makecell}
\usepackage{pifont}
\usepackage[normalem]{ulem}
\useunder{\uline}{\ul}{}

\def\BibTeX{{\rm B\kern-.05em{\sc i\kern-.025em b}\kern-.08em
    T\kern-.1667em\lower.7ex\hbox{E}\kern-.125emX}}
\begin{document}

\title{Expressive Speech Retrieval using Natural Language Descriptions of Speaking Style}

\author{
\IEEEauthorblockN{Wonjune Kang}
\IEEEauthorblockA{\textit{Massachusetts Institute of Technology} \\
Cambridge, MA, USA}
\and
\IEEEauthorblockN{Deb Roy}
\IEEEauthorblockA{\textit{Massachusetts Institute of Technology} \\
Cambridge, MA, USA}
}

\maketitle

\begin{abstract}
    We introduce the task of \textit{expressive speech retrieval}, where the goal is to retrieve speech utterances spoken in a given style based on a natural language description of that style. While prior work has primarily focused on performing speech retrieval based on \textit{what} was said in an utterance, we aim to do so based on \textit{how} something was said. We train speech and text encoders to embed speech and text descriptions of speaking styles into a joint latent space, which enables using free-form text prompts describing emotions or styles as queries to retrieve matching expressive speech segments. We perform detailed analyses of various aspects of our proposed framework, including encoder architectures, training criteria for effective cross-modal alignment, and prompt augmentation for improved generalization to arbitrary text queries. Experiments on multiple datasets encompassing 22 speaking styles demonstrate that our approach achieves strong retrieval performance as measured by Recall@$k$.  
\end{abstract}

\begin{IEEEkeywords}
expressive speech retrieval, spoken content retrieval, cross-modal retrieval, style representation learning, contrastive learning
\end{IEEEkeywords}
\section{Introduction}
\label{section:introduction}

Speech retrieval, or spoken content retrieval, is the task of indexing and retrieving spoken content based on audio rather than text descriptions or other metadata~\cite{chelba2008retrieval, lee2009voice, larson2012spoken, lee2015spoken}.
Typically, users submit a query for the content they are searching for, and the retrieval system searches over a database of spoken audio segments to return relevant results.
The task has become increasingly important in the era of Internet-scale multimedia as people navigate vast quantities of spoken data online.
More recently, speech retrieval has also seen interest in the context of retrieval-augmented generation (RAG) for large language models (LLMs)~\cite{lin2024speechdpr, wang2024retrieval, chen2025wavrag, min2025speech}, as it can enable LLMs to access and leverage additional context from spoken documents.

Most prior work in speech retrieval has focused solely on semantic content; that is, retrieving speech based on \textit{what} was said, rather than \textit{how} it was said.
However, spoken language contains rich paralinguistic features, which can carry additional information for interpretation and downstream use.
In the context of retrieval, enabling queries that target expressive aspects of speech could significantly broaden the utility of such systems.
For example, it could enable retrieving emotionally significant moments from political debates, extracting enthusiastic clips from podcasts for highlight generation, or filtering for calm and empathetic tones in therapy or customer service audio.

In this paper, we introduce the task of \textit{expressive speech retrieval}, where the objective is to retrieve speech segments that are spoken in a given emotion or speaking style, as described by a text prompt provided by a user.
While prior work has explored emotional audio retrieval using a limited set of predefined emotion classes~\cite{dhamyal2024prompting}, we extend this paradigm by generalizing beyond discrete emotions to a broader range of speaking styles.
This includes not only emotional categories such as ``angry'' or ``sad'', but also styles such as ``bored'', ``confused'', ``laughing'', and ``sarcastic''.
Furthermore, we enable the use of general, free-form text prompts for querying and retrieving content in an open-ended setting, which was not addressed previously.

Our system consists of two components: a speech encoder that processes input audio, and a text encoder that processes natural language prompts describing speaking styles.
The training framework aims to learn cross-modal speech and text representations that are aligned within a shared latent space, which enables retrieval at inference time by comparing text query embeddings with speech embeddings based on cosine similarity.
To achieve this, we train the two encoders using a contrastive learning framework based on Contrastive Language-Audio Pretraining (CLAP)~\cite{wu2023large}.
We also utilize an adversarial loss by applying a modality discriminator to the speech and text embeddings, which encourages the encoders to produce representations that are more tightly aligned in the latent space.
To determine the most effective model configurations, we additionally explore various speech representation models and language models as backbones for the encoders.

Furthermore, to improve generalization to diverse user queries, we augment the training data with stylistic prompt variation.
Starting from a base set of 22 speaking styles from 3 datasets (IEMOCAP~\cite{busso2008iemocap}, ESD~\cite{zhou2022emotional}, and Expresso~\cite{nguyen2023expresso}), we use an LLM to generate multiple paraphrased and stylistically varied prompts for each category.
This prompt augmentation encourages our method to handle a wider variety of stylistic descriptions and reduce its reliance on fixed label sets.

Experiments demonstrate that our proposed framework is able to reliably retrieve speech segments that match queried emotions and styles across a broad range of prompts.
Our findings contribute to a growing line of work that aims to bridge text and expressive speech in machine learning systems, and we believe they also offer valuable implications for related tasks such as prompt-conditioned speech synthesis~\cite{guo2023prompttts, liu2023promptstyle, leng2024prompttts2, hai2024dreamvoice} and speech style captioning~\cite{yamauchi2024stylecap, kawamura2024librittsp}.\footnote{Code, data, and model checkpoints are available at \texttt{\label{footnote:code}\url{https://github.com/wonjune-kang/expressive-speech-retrieval}}.}
\begin{figure*}[t]
    \centering
    \includegraphics[height=2.6in]{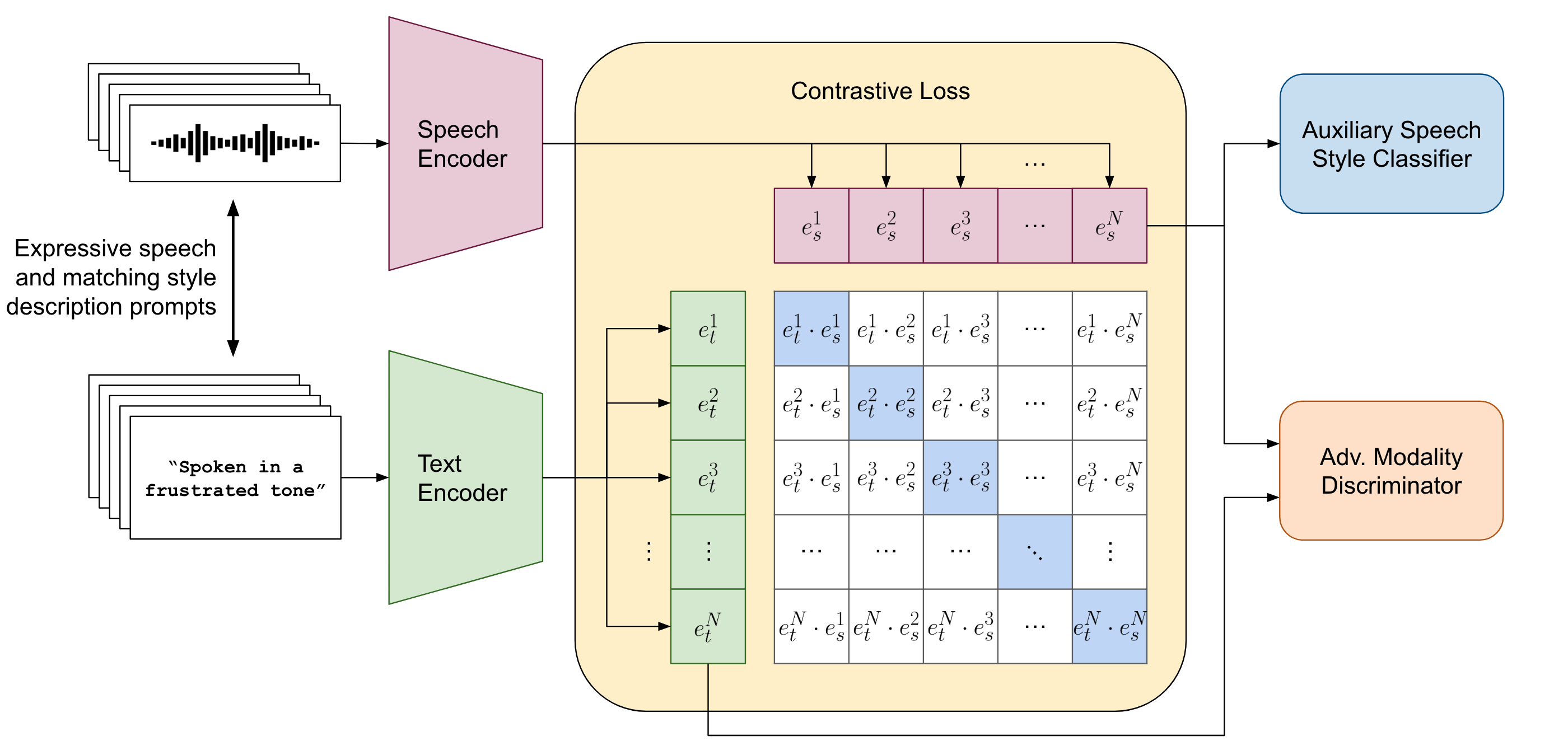}
    \vspace{3pt}
    \caption{Proposed training framework for expressive speech retrieval. Using a contrastive loss, speech and matching text prompts describing speaking style are projected into a joint embedding space. An adversarial modality discriminator with a gradient reversal layer further encourages alignment, and an auxiliary style classification loss is applied to speech embeddings.}
    \label{fig:training}
    \vspace{-3pt}
\end{figure*}

\section{Background and Related Work}
\label{section:background}

\subsection{Speech retrieval}

Traditionally, speech retrieval has been treated as an extension of text retrieval by transcribing audio using automatic speech recognition (ASR), followed by performing text-based keyword or semantic search over the transcribed content~\cite{mamou2007vocabulary, chia2010statistical, lee2013improved}.
However, these cascaded approaches are susceptible to transcription errors which can propagate and degrade performance.
To address this, prior work explored optimizing ASR systems specifically for retrieval tasks~\cite{van2007evaluating, larson2009investigating, he2013speech} or eliminated the transcription step altogether by indexing and retrieving directly in the audio domain~\cite{chan2010unsupervised, anguera2013memory, metze2013spoken}.

More recently, speech retrieval has seen increasing interest in the context of speech-based retrieval-augmented generation (RAG) for large language models (LLMs)~\cite{lin2024speechdpr, wang2024retrieval, chen2025wavrag, min2025speech}.
In this setting, some works have used contrastive learning to train speech embedding models that can match spoken or text queries to audio passages~\cite{wang2024retrieval, ghosh2024recap}, while others have used distillation methods to train speech retrievers from text retrievers~\cite{lin2024speechdpr, min2025speech}.
These approaches have been applied to improve LLMs' performance on tasks such as ASR for named entities~\cite{wang2024retrieval} or spoken question answering~\cite{lin2024speechdpr, min2025speech}.

The aforementioned methods focus exclusively on retrieval based on the semantic content of speech utterances.
In contrast, our work explores retrieval of expressive speech based on descriptions of paralinguistics: specifically, emotions and speaking style.
This task represents an approach to speech retrieval that has been underexplored thus far.

\subsection{Aligning speech with text descriptions of style}

A growing body of work has studied the alignment of expressive speech with natural language descriptions of style.
One popular application is prompt-guided speech synthesis, including text-to-speech (TTS)~\cite{guo2023prompttts, liu2023promptstyle, leng2024prompttts2} and voice conversion~\cite{hai2024dreamvoice}, which allows users to control the prosody and expressiveness of synthesized speech through natural language prompts.
Another related task is speech style captioning, which involves generating textual descriptions of speaking style from speech inputs~\cite{yamauchi2024stylecap, kawamura2024librittsp}.
Recently, \cite{huo2023beyond} introduced a method for target speech extraction that uses natural language descriptions of speaking style as a conditioning feature to extract desired speech from a mixture of sound sources.

The work perhaps most closely related to ours is \cite{dhamyal2024prompting}, which proposed a contrastive learning approach for aligning emotion-related words with emotional speech.
Their representation learning method was evaluated on emotional audio retrieval and speech emotion recognition.
However, this work limited its scope to a small set of emotional classes, and it only considered a few types of prompts that were centered around the class label itself.
We extend this paradigm by going beyond just emotion categories and covering a broader range of speaking styles.
Furthermore, we utilize a much wider range of natural language prompts for describing each style during training, enabling more flexibility and generalization to diverse user queries.
To the best of our knowledge, our work is the first to address expressive speech retrieval using natural language descriptions of speaking style in a general, open-ended setting.
\section{Retrieving Speech with Text Descriptions of Speaking Style}
\label{section:proposed_method}

Our framework for expressive speech retrieval consists of two encoders, one for speech and one for text.
The speech encoder processes raw audio inputs, while the text encoder takes in natural language prompts describing emotion or speaking style.
The two encoders are trained using a contrastive learning objective that aligns their outputs in a joint latent space of dimension $d$, which enables cross-modal embedding comparisons using cosine similarity.
An overview of our proposed training framework is shown in Fig.~\ref{fig:training}.

\subsection{Speech encoder}

We consider two speech representation models as the backbone for our speech encoder: WavLM~\cite{chen2022wavlm} and emotion2vec~\cite{ma2024emotion2vec}.
WavLM is a general-purpose speech representation model that has demonstrated strong performance across a wide range of speech processing tasks.
Meanwhile, emotion2vec is a model specifically designed for speech emotion representation learning and related tasks, such as speech emotion recognition, song emotion recognition, and sentiment analysis.
While the backbone models are fine-tuned during training, we aim to determine the extent to which starting out with a paralinguistics-specific representation model like emotion2vec might offer advantages over a general purpose model like WavLM for our task.

In our experiments, we use the WavLM Base+ and emotion2vec-base configurations of the models, each of which contains $\sim$94M parameters.
We temporally mean pool the output embedding sequence from the backbone model and feed the resulting vector through a linear projection layer to produce the final $d$-dimensional speech embedding $e_s \in \mathbb{R}^d$.

\subsection{Text encoder}

We similarly explore several pretrained language models as backbones for the text encoder: BERT~\cite{devlin2019bert}, RoBERTa~\cite{liu2019roberta}, T5~\cite{raffel2020exploring}, and Flan-T5~\cite{longpre2023flan}.
For the T5-based models, we use only the encoder component of the full encoder-decoder architecture.
All models are evaluated in their base configurations, with BERT-base, T5-base, and Flan-T5-base containing $\sim$110M parameters, and RoBERTa-base containing $\sim$125M parameters.

For BERT and RoBERTa, we take the embedding of the \texttt{[CLS]} token as the prompt-level representation.
For T5 and Flan-T5, we follow \cite{ni2022sentence} and apply mean pooling over the model outputs to obtain the full prompt representation.
As with the speech encoder, a linear projection layer is applied to map the output of the encoder backbone to a $d$-dimensional text embedding $e_t \in \mathbb{R}^d$.

\subsection{Training}

\subsubsection{Contrastive learning}

To align speech and text embeddings in a shared latent space, we adopt a contrastive learning objective based on CLAP~\cite{wu2023large}.
Given a batch of $N$ paired speech and text samples $\{(x_s^i, x_t^i)\}_{i=1}^N$, we encode each speech input $x_s^i$ and text prompt $x_t^i$ into embeddings $e_s^i = E_s(x_s^i)$ and $e_t^i = E_t(x_t^i)$, where $E_s$ and $E_t$ denote the speech and text encoders respectively, including the final projection layers.

The embeddings are $L_2$-normalized to unit length and compared using cosine similarity.
The contrastive loss encourages matched pairs $(e_s^i, e_t^i)$ to have higher similarity than all unmatched pairs $(e_s^i, e_t^j)$ where $j \neq i$.
We use a symmetric loss as follows:
\begin{equation}
    \mathcal{L}_{\text{contrast}} = \frac{1}{2N} \sum_{i=1}^N \left[ \ell(e_s^i, e_t^i) + \ell(e_t^i, e_s^i) \right]
\end{equation}
where the individual loss terms are computed as:
\begin{equation}
    \ell(e_a^i, e_b^i) = -\log \frac{\exp(\text{sim}(e_a^i, e_b^i)/\tau)}{\sum_{j=1}^N \exp(\text{sim}(e_a^i, e_b^j)/\tau)}.
\end{equation}
Here, $\text{sim}(\cdot, \cdot)$ denotes cosine similarity and $\tau$ is a learnable temperature parameter that scales the range of logits.

\vspace{2pt}
\subsubsection{Adversarial modality discriminator}

We utilize an adversarial modality discriminator~\cite{ganin2016domain} to reduce modality-specific variation and encourage modality-invariant representations.
The discriminator $D$ is a two-layer feedforward neural network with hidden layer dimension $d_D$ and a ReLU activation.
It takes speech and text embeddings as input and is trained simultaneously with the two encoders to predict which modality an embedding is computed from.
The discriminator network incorporates a gradient reversal layer (GRL)~\cite{ganin2015unsupervised} to perform adversarial training on the encoders.

Formally, the discriminator minimizes the binary cross-entropy loss for modality classification:
\begin{gather}
    \mathcal{L}_{\text{disc}} = - \mathbb{E}_{e_s \sim \mathcal{E}_s} \log D(e_s) - \mathbb{E}_{e_t \sim \mathcal{E}_t} \log (1 - D(e_t)), \\
    \mathcal{L}_{\text{adv}} = -\mathcal{L}_{\text{disc}},
\end{gather}
where $\mathcal{E}_s$ and $\mathcal{E}_t$ are the sets of speech and text embeddings, respectively.
Meanwhile, $\mathcal{L}_{\text{adv}}$ is applied to the encoders, as they are trained to maximize $\mathcal{L}_{\text{disc}}$ via the GRL.
This encourages the learned embeddings to be more indistinguishable across modalities and leads to tighter alignment between speech and text embeddings in the shared latent space.

\vspace{2pt}
\subsubsection{Auxiliary style classification loss}

To ensure that the speech encoder effectively captures style-related information relevant for retrieval, we incorporate an auxiliary style classification objective, $\mathcal{L}_{\text{cls}}$.
Specifically, we use a two-layer feedforward neural network $C$ with hidden layer dimension $d_C$ and a ReLU activation; the network takes speech embeddings as input and is trained to predict the emotion or style label of each utterance using cross-entropy loss.
This loss encourages the speech encoder to produce embeddings that contain more discriminative information with respect to style; we found it to be critical for learning speech embeddings that could successfully align with the text embeddings.\footnote{However, we found that the text encoder did not require a similar loss.}

\vspace{2pt}
\subsubsection{Overall training objective}

The final training objective for the encoders is a weighted sum of the above loss terms:
\begin{equation}
    \mathcal{L}_{\text{total}} = \lambda_{\text{contrast}} \mathcal{L}_{\text{contrast}} + \lambda_{\text{adv}} \mathcal{L}_{\text{adv}} + \lambda_{\text{cls}} \mathcal{L}_{\text{cls}},
\end{equation}
where the weights $\lambda_{\text{contrast}}$, $\lambda_{\text{adv}}$, and $\lambda_{\text{cls}}$ balance the relative contributions of each loss.

\subsection{Inference time retrieval}

\begin{figure}[t]
    \centering
    \includegraphics[width=1.0\columnwidth]{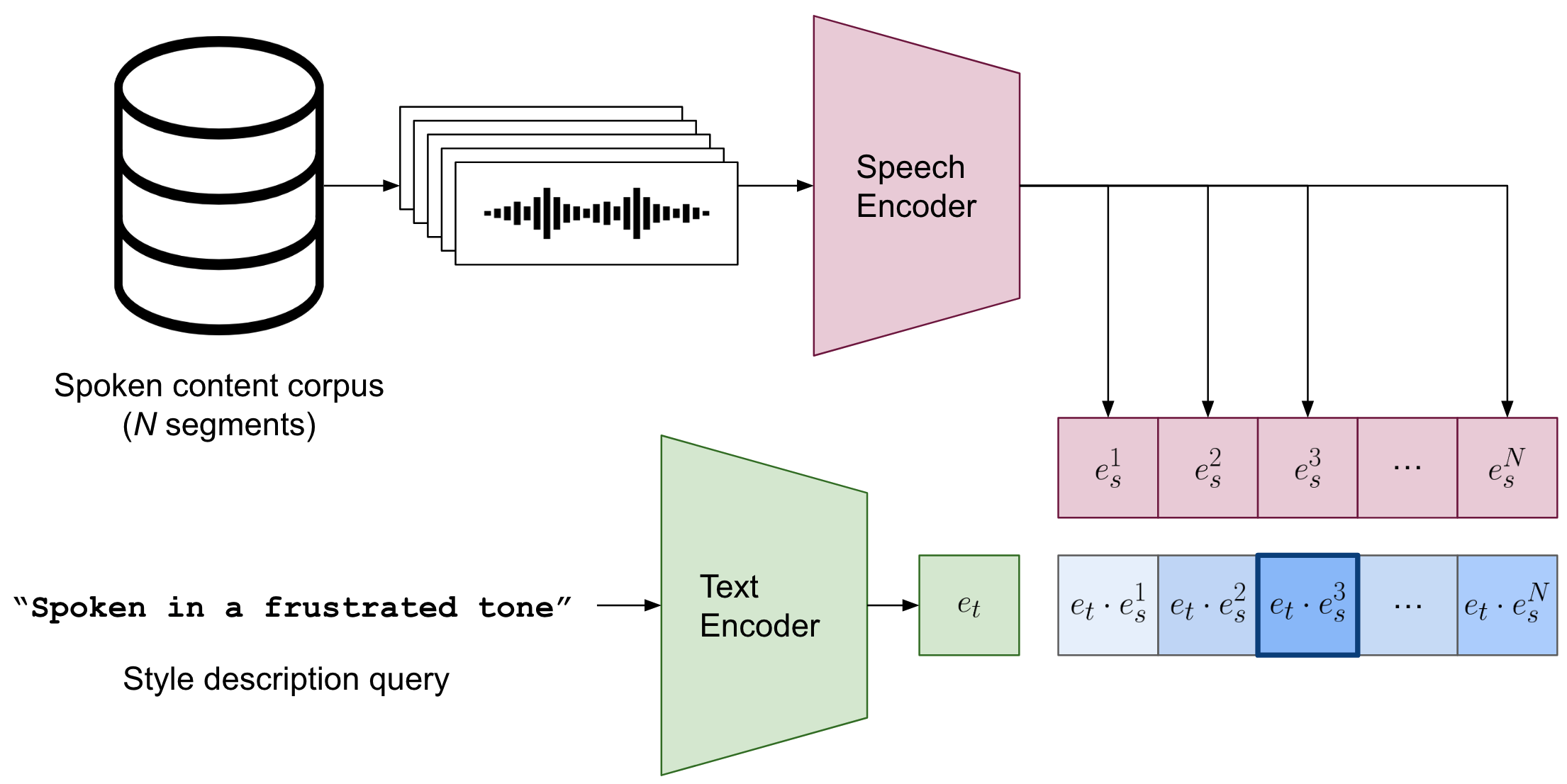}
    \caption{Expressive speech retrieval at inference time. Given a natural language prompt describing a speaking style, the text encoder produces a query embedding which is used to retrieve matching utterances from a speech corpus.}
    \label{fig:retrieval}
    \vspace{-8pt}
\end{figure}

At inference time, the speech encoder is used to compute and cache speech style embeddings for all utterances in a target speech corpus.
Then, given a user's natural language prompt describing a desired speaking style, the text encoder is used to generate a text embedding which behaves as a query.
Retrieval can be performed by computing the cosine similarity between the text embedding and all cached speech embeddings, and returning the most similar utterances, optionally thresholding based on similarity score.
Fig.~\ref{fig:retrieval} illustrates the retrieval process.

\section{Experimental Setup}
\label{section:experiments}

\vspace{-2pt}
\subsection{Data}
\label{subsec:data}
\vspace{-1pt}

\subsubsection{Datasets}

We conduct experiments using three datasets: IEMOCAP~\cite{busso2008iemocap}, ESD~\cite{zhou2022emotional}, and Expresso~\cite{nguyen2023expresso}.
IEMOCAP consists of speech from actors performing scripted and improvised scenarios designed to elicit emotional expression.
We use data corresponding to 9 of the 10 annotated emotion classes, excluding the ``other'' category.
ESD is an emotional speech dataset consisting of Chinese and English speech covering 5 emotions.
For our experiments, we use only the English portion of the dataset.
Expresso is an expressive speech corpus containing read and improvised speech across 26 annotated style categories.
These include not only emotions, but also a broader range of speaking styles such as ``bored'', ``confused'', ``laughing'', and ``sarcastic''.
We filter out certain categories that we deem less relevant for our task (e.g., ``animal directed'', ``child directed'', ``narration''), utilizing 19 of the 26 styles.

In total, our data encompass approximately 64 hours of speech spanning a final set of 22 unique emotion and style classes: \textit{angry, awe, bored, calm, confused, desire, disgusted, enunciated, excited, fast, fearful, frustrated, happy, laughing, neutral, projected, sad, sarcastic, sleepy, surprised, sympathetic, whispered}.
All audio was resampled to 16kHz.
We use the entirety of IEMOCAP for training, while we use the standard train/validation/test splits for ESD and Expresso.
Full dataset statistics as used in our experiments are shown in Table~\ref{table:dataset_statistics}.

\subsubsection{Prompt generation}

To improve our system's generalizability to diverse text prompts, we use a large language model, GPT-4o~\cite{hurst2024gpt}, to generate 11 natural language prompt templates for describing speaking style.
For each of the 22 style classes, we also generate 5 paraphrased synonyms or descriptive phrases that can be substituted into these templates.
During training, we dynamically sample a template and synonym based on an utterance's class label to construct a random style description.\footnote{The full list of prompt templates and synonyms used in our experiments is included in our open-sourced codebase; see footnote~\ref{footnote:code}.}

\vspace{-1pt}
\subsection{Training configurations}
\vspace{-1pt}

We use a batch size of $N = 32$ for the contrastive learning objective and initialize $\tau$ to 0.07.
The dimension of the cross-modal latent space $d$ is set to 512. The modality discriminator and auxiliary style classifier use a hidden layer of size 128 ($d_D = 128$, $d_C = 128$).
The loss weights are set to $\lambda_\text{contrast} = 1.0$, $\lambda_\text{adv} = 0.1$, and $\lambda_\text{cls} = 0.5$.
We use the AdamW optimizer~\cite{loshchilov2019decoupled} with a learning rate of 5e-5 and $\beta_1 = 0.9, \beta_2 = 0.999$ for the speech and text encoders, as well as for the auxiliary style classifier.
For the modality discriminator, we use AdamW with a reduced learning rate of 2e-5 and the same $\beta$ parameters.

We fine-tune the entire speech encoder during training, while for the text encoder, we freeze the pre-trained language model backbone and update only the linear projection layer.
We also use a multi-stage training schedule.
First, we pre-train the speech encoder using only the auxiliary style classification loss for 5 epochs.
Then, we introduce the contrastive loss and train with both objectives for another 5 epochs, after which we begin incorporating the adversarial loss.
We found this strategy to be helpful for stabilizing training and allowing the encoders to effectively learn and align their cross-modal representations.

All models are trained on a single NVIDIA A6000 Ada 48GB GPU using automatic mixed precision with bfloat16.
We use a maximum audio segment length of 8 seconds per batch.
Training is performed for up to 110 epochs (around 171k steps), with evaluation checkpoints selected based on validation loss.

\begin{table}[t]
    \centering
    \caption{Statistics of the datasets used in our experiments.}
    \begin{tabularx}{0.95\linewidth}{l c *3{>{\centering\arraybackslash}X}}
    \toprule
    \multirow{2}{*}{\vspace{-6pt}\textbf{Dataset}} &
    \multirow{2}{*}{\textbf{\begin{tabular}[c]{@{}c@{}} \# of used \\ \vspace{-5pt}emotions/styles\end{tabular}}} &
    \multicolumn{3}{c}{\textbf{Amount of data (hrs)}} \\ \cmidrule(lr){3-5} & & 
    train & val & test \\ \midrule
    IEMOCAP~\cite{busso2008iemocap}     & 9     & 12.4           & 0.0          & 0.0           \\
    ESD~\cite{zhou2022emotional}        & 5     & 11.4           & 0.8          & 1.3           \\
    Expresso~\cite{nguyen2023expresso}  & 19    & 36.9           & 0.6          & 0.6           \\ \midrule
    Total                               & 22    & 60.7           & 1.4          & 1.9           \\ \bottomrule
    \end{tabularx}
    \label{table:dataset_statistics}
    \vspace{-2pt}
\end{table}

\begin{table*}[t]
    \centering
    \caption{Expressive speech retrieval results on the ESD and Expresso datasets. \textbf{Bold} and \underline{underlined} values indicate the best and second best scores for non-baseline models, respectively. Note that the baselines are limited to classification (and thus, retrieval) over only the fixed set of styles that they were trained on. In contrast, our approach is able to support a wider range of free-form natural language queries, which are not limited to the discrete style classes seen during training.}
    \vspace{-2pt}
    \begin{tabular}{lcccccccc}
    \toprule
    \multirow{2}{*}{\vspace{-7pt}\textbf{Model}} &
    \multicolumn{4}{c}{\textbf{ESD}}    &
    \multicolumn{4}{c}{\textbf{Expresso}}   \\
    \cmidrule(lr){2-5}
    \cmidrule(lr){6-9}
                                         & \textbf{Recall@1}        & \textbf{Recall@5}        & \textbf{Recall@10}       & \textbf{Recall@20}       & \textbf{Recall@1}        & \textbf{Recall@5}        & \textbf{Recall@10}       & \textbf{Recall@20}       \\ \midrule
    Baseline: WavLM classifier & 0.6106          & 0.9270          & 0.9562          & 0.9792          & 0.8144          & 0.9399          & 0.9650          & 0.9820          \\
    Baseline: emotion2vec classifier   & 0.6814          & 0.9288          & 0.9582          & 0.9764          & 0.9467          & 0.9784          & 0.9821          & 0.9894          \\ \midrule
    BERT + WavLM                & 0.3350          & 0.8400          & 0.9394          & 0.9710          & 0.3623          & 0.9013          & 0.9631          & 0.9720          \\
    RoBERTa + WavLM             & 0.4830          & 0.9024          & 0.9484          & 0.9610          & 0.6284          & 0.8640          & 0.9397          & 0.9616          \\
    T5 + WavLM                  & 0.4100          & 0.8718          & 0.9506          & 0.9714          & 0.4846          & 0.8677          & 0.9511          & 0.9616          \\
    Flan-T5 + WavLM             & 0.3326          & 0.8974          & 0.9544          & 0.9660          & 0.6104          & 0.8826          & 0.9526          & 0.9619          \\ \midrule
    BERT + emotion2vec          & 0.3414          & 0.8726          & 0.9600          & {\ul 0.9738}    & \textbf{0.8317} & 0.9773          & 0.9846          & 0.9846          \\
    RoBERTa + emotion2vec       & {\ul 0.6026}    & \textbf{0.9338} & \textbf{0.9660} & \textbf{0.9760} & 0.8159          & \textbf{0.9826} & {\ul 0.9857}    & \textbf{0.9894} \\
    T5 + emotion2vec            & \textbf{0.6438} & 0.9114          & 0.9494          & 0.9666          & 0.7617          & 0.9793          & 0.9830          & 0.9853          \\
    Flan-T5 + emotion2vec       & 0.3864          & {\ul 0.9232}    & {\ul 0.9618}    & 0.9730          & {\ul 0.8284}    & {\ul 0.9806}    & \textbf{0.9860} & {\ul 0.9873}    \\ \bottomrule
    \end{tabular}
    \label{table:main_results}
    \vspace{-6pt}
\end{table*}

\subsection{Evaluation}

We adopt a controlled protocol to evaluate the retrieval performance of the various models.
For each style class, we randomly construct a set of 1000 ``retrieval trials'', where each trial is generated by randomly sampling one speech segment that matches the given target class (a positive sample) and 400 distractor utterances sampled from all other classes (negative samples).
In other words, each trial is set up to have exactly one relevant target within a set of 401 candidates.

To perform evaluation for a given trial, we consider all possible combinations of prompt templates and synonyms for the target style as potential queries and compute their text embeddings.
Then, we compute the cosine similarities between all of the potential text query embeddings and the speech embeddings of the 401 candidate utterances in the trial.
The candidate utterances are ranked by their average similarity across all possible query prompts.
We measure performance using Recall@$k$ for $k \in \{1, 5, 10, 20\}$, where a model is considered to have successfully performed retrieval at rank $k$ if the target (positive) segment appears within the top $k$ of the ranked candidates.
Final results are averaged over all 1000 trials for each style.

\subsection{Baselines}

To the best of our knowledge, no prior work has addressed the task of expressive speech retrieval as framed in this paper.
While the model used for emotional audio retrieval in \cite{dhamyal2024prompting} is structurally similar to our system, it lacks several components of ours, including the modality discriminator, auxiliary style classification loss, and prompt augmentation.
(We explore the impact of these additions in Section~\ref{subsec:ablations}.)

Therefore, we compare our proposed framework with a classification-based retrieval baseline.
Specifically, we fine-tune WavLM and emotion2vec as style classifiers on our expressive speech datasets; they achieve 93.06\% and 94.66\% classification accuracy on the combined ESD and Expresso test sets, respectively.
Then, to perform retrieval for a target style, we compute the class prediction logits for each sample in a given evaluation trial.
We rank the samples by their logit values for the target class and evaluate performance using Recall@$k$.
\section{Results}
\label{section:results}

\subsection{Retrieval performance}

Table~\ref{table:main_results} shows the main expressive speech retrieval results from our experiments.
Our best-performing model configurations (RoBERTa + emotion2vec and Flan-T5 + emotion2vec) overall achieve performance on par with, or even slightly exceeding, that of the classification-based baselines.
However, note that the baselines are limited to classifying (and therefore retrieving) only speech based on classes that they were trained on.
Meanwhile, our proposed framework enables retrieval based on arbitrary natural language queries, which need not perfectly match the style classes seen during training.
This offers significantly more flexibility for real-world applications.

For models trained using our approach, we find that architecture choices have a clear impact on performance.
For the speech encoder, model configurations using emotion2vec consistently outperform those using WavLM across both datasets.
This is particularly evident on Expresso, which contains styles like ``confused'' that may require more nuanced paralinguistic features to process and represent accurately.
Our results suggest that starting out with an emotion-specific speech representation model enables fine-grained information to be captured more effectively, which in turn leads to better downstream performance.

\begin{figure}[t]
    \centering
    \includegraphics[width=0.9\columnwidth]{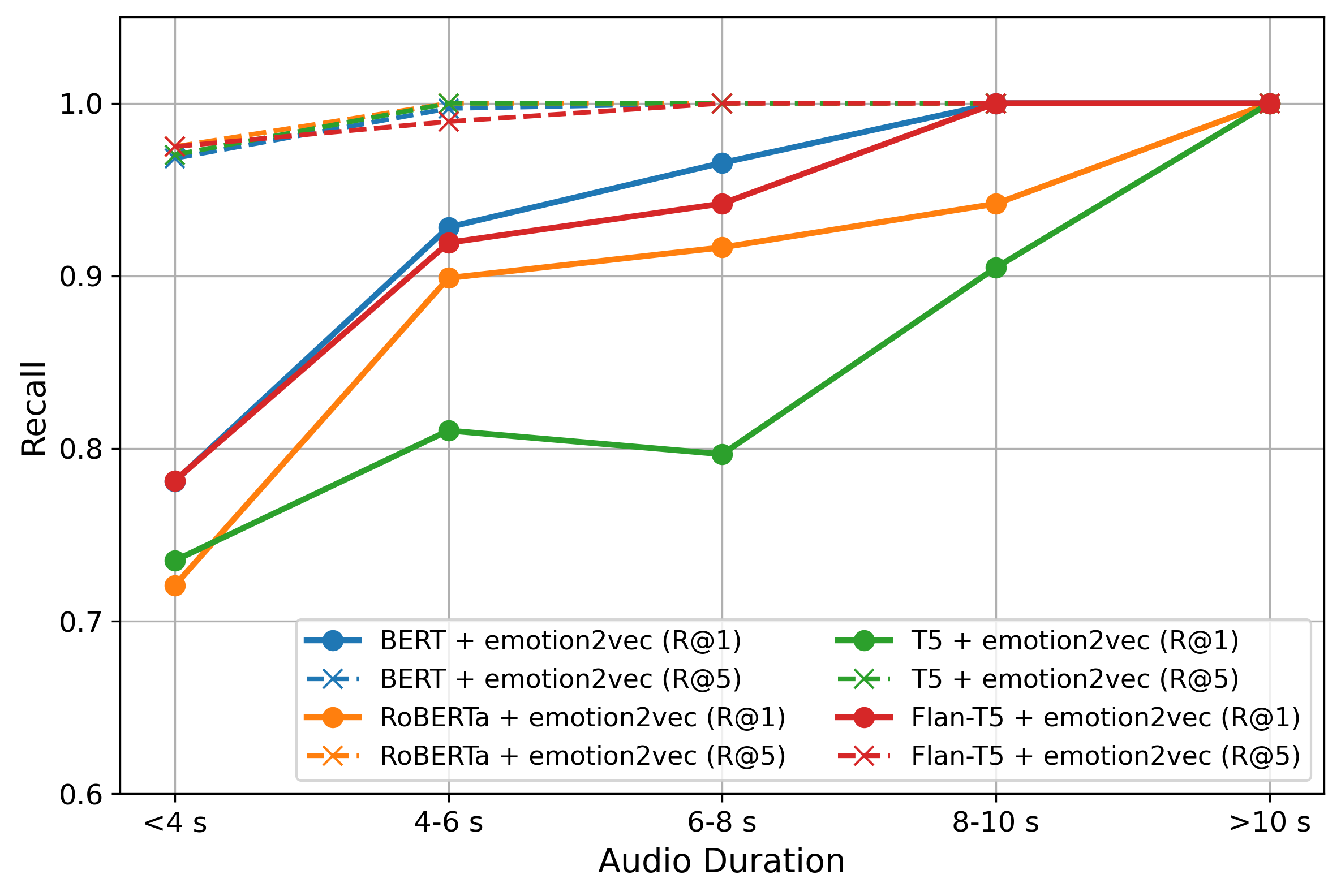}
    \vspace{-4pt}
    \caption{Retrieval performance on Expresso (Recall@1 and Recall@5) as a function of the duration of the target speech to be retrieved, bucketed into bins.}
    \label{fig:performance_audio_length}
    \vspace{-6pt}
\end{figure}

On the text encoder side, we observe that using stronger language models provides better retrieval performance, with RoBERTa and Flan-T5 consistently outperforming BERT and T5 across both datasets.
While these results may intuitively make sense, prior works that used pre-trained language models to encode text prompts for speech applications largely overlooked this aspect of model design, often defaulting to BERT without testing alternatives~\cite{dhamyal2024prompting, liu2023promptstyle, guo2023prompttts, leng2024prompttts2, shimizu2024prompttts++, lee2024promoticon}.
Our findings demonstrate that even for encoding relatively short style descriptions, the capacity of the language model has an impact on downstream performance.

\vspace{-2pt}
\subsection{Performance across target utterance lengths}
\vspace{-2pt}

To better understand the capabilities of our system, we analyze performance as a function of target utterance duration.
We group utterances in the Expresso test set into five duration bins (<4 s, 4-6 s, 6-8 s, 8-10 s, and >10 s) and compute Recall@$k$ for each bin.
Fig.~\ref{fig:performance_audio_length} shows scores for the models using emotion2vec as the speech encoder for $k = 1$ and $k = 5$.

We find that retrieval performance is sensitive to the length of the target segment, especially at Recall@1; for all model configurations, we see a clear trend where longer utterances yield better retrieval accuracy.
This aligns with intuition, since longer speech utterances are likely to contain more acoustic information that determines expressive style.
However, we find that performance across all audio length bins is quite strong at Recall@5, with all models notably achieving perfect retrieval performance for utterances longer than 6 seconds.

Overall, these results demonstrate the strength of our proposed framework: it effectively retrieves expressive speech across a range of target speech durations, and is especially reliable when allowed a small margin for ambiguity in shorter utterances (e.g., top-5 results).

\begin{table}[t]
    \centering
    \caption{Retrieval results for various ablations on the RoBERTa + emotion2vec model on the Expresso test set.}
    \vspace{-2pt}
    \begin{tabular}{lcccc}
    \toprule
    \multirow{2}{*}{\vspace{-6pt}\textbf{Model}} &
    \multicolumn{4}{c}{\textbf{Expresso}}  \\
    \cmidrule(lr){2-5}
    & \textbf{R@1}  & \textbf{R@5}  & \textbf{R@10} & \textbf{R@20}    \\ \midrule
    RoBERTa + emotion2vec               & 0.8159    & 0.9826   & 0.9857 & 0.9894    \\ \midrule
    w/o pre-training speech encoder     & 0.7789    & 0.9196   & 0.9803 & 0.9811    \\
    w/o aux. style classification loss  & --        & --       & --     & --        \\
    w/o modality discriminator          & 0.6714    & 0.8653   & 0.9733 & 0.9749    \\
    w/o prompt augmentation             & 0.5901    & 0.8487   & 0.9683 & 0.9761    \\
    \bottomrule
    \end{tabular}
    \label{table:ablation_results}
    \vspace{-5pt}
\end{table}

\vspace{-1pt}
\subsection{Ablation studies}
\label{subsec:ablations}
\vspace{-1pt}

We conduct ablation studies to assess the impact of certain key components of our training framework using the best-performing model configuration (RoBERTa + emotion2vec).
Specifically, we examine the impact of pre-training the speech encoder, the auxiliary style classification loss, the adversarial modality discriminator, and prompt augmentation.
For ablating prompt augmentation, we train using a single fixed prompt template (``Speech that is <style>.'').
Then, we only use the style class label itself in this prompt during training, without using any other synonyms.
Results are shown in Table~\ref{table:ablation_results}.

Pre-training the speech encoder on style classification before introducing the remaining training objectives leads to modest but consistent performance improvements.
Meanwhile, removing the auxiliary style classification loss altogether results in the model completely failing to learn a meaningful joint embedding space, which causes retrieval performance to collapse.
These results suggest that some degree of explicit supervision is crucial to facilitate cross-modal learning and encourage the speech encoder to effectively capture style-related information, even if the pre-trained backbone model may already represent some paralinguistic information.

\begin{figure}[t]
    \centering
    \includegraphics[width=0.97\columnwidth]{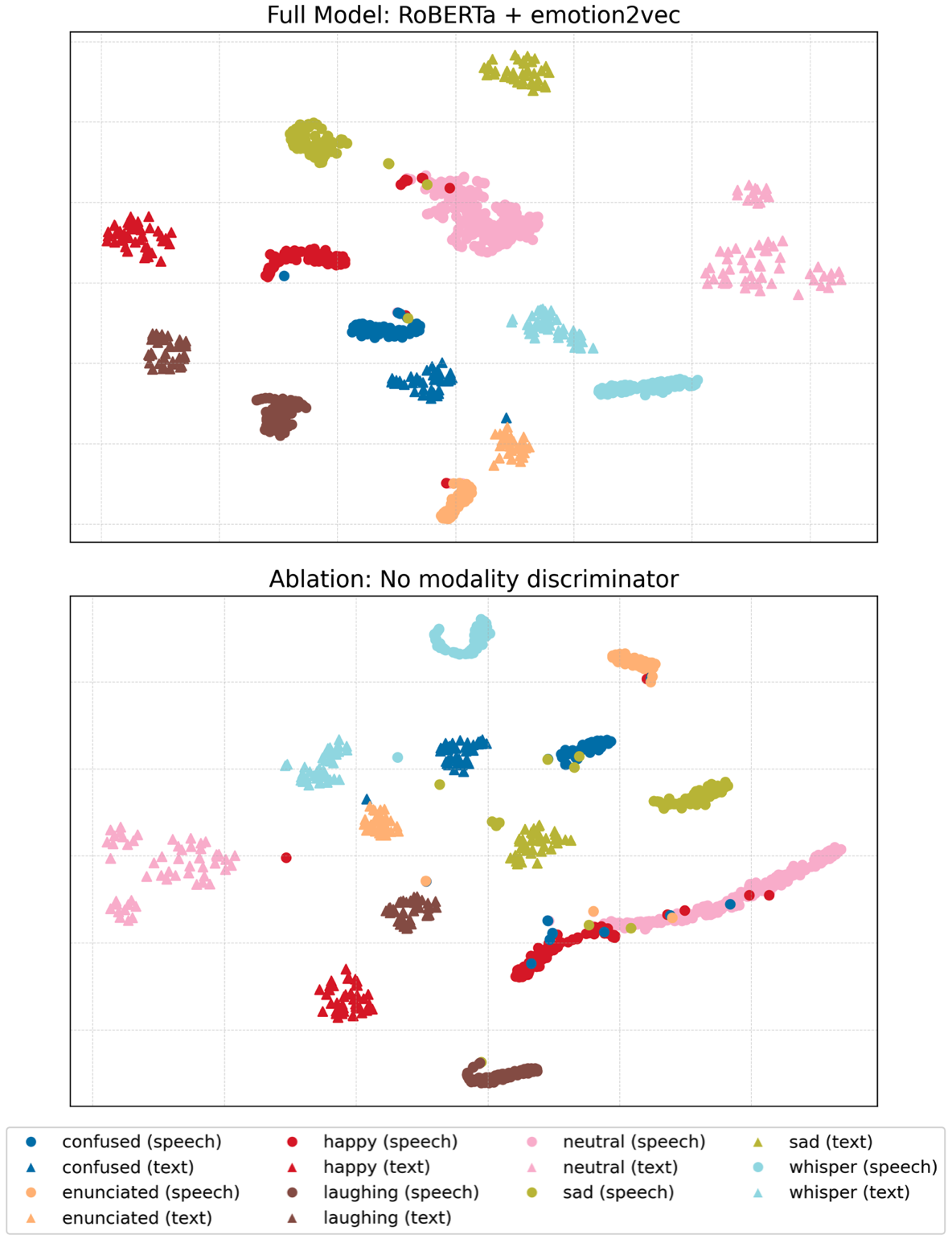}
    \caption{t-SNE plots of speech and text embeddings computed on Expresso test set samples using the full RoBERTa + emotion2vec model (top) and the ablation trained without the modality discriminator (bottom).}
    \label{fig:tsne}
    \vspace{-6pt}
\end{figure}

Removing the modality discriminator leads to a significant drop in performance.
This demonstrates the importance of adversarial training for aligning the speech and text modalities; the discriminator encourages the encoders to produce embeddings that are more tightly aligned than what contrastive loss alone can achieve.
This is visualized using t-SNE plots~\cite{van2008visualizing} in Figure~\ref{fig:tsne}.
Note that in the full model, speech and text embeddings are largely overlapping in the shared latent space.
In the ablation, they are less well aligned and clearly occupy different areas of the latent space, with speech embeddings in the right side of the plot and text embeddings in the left.

Finally, omitting prompt augmentation significantly degrades performance when performing retrieval using a more diverse set of stylistic descriptions as prompts.
This confirms that training with a wide variety of prompts is essential for enabling the system to generalize to the numerous kinds of user queries that may arise in real-world situations.

\section{Conclusion}
\label{section:conclusion}

We presented a framework for expressive speech retrieval that enables using free-form text descriptions of style as queries, covering a broad range of general speaking styles.
By training speech and text encoders to learn a joint latent space between paralinguistic speech features and language, our framework effectively enables retrieval via similarity-based comparisons between text query embeddings and speech embeddings.
To determine the optimal approach for addressing the task, we performed detailed analyses of various components of our proposed framework, including encoder architectures, training criteria for cross-modal alignment, and prompt augmentation for generalizing to more diverse queries.
Beyond speech retrieval, we believe our findings also offer noteworthy insights for other tasks that link natural language prompting with speech.

\clearpage

\bibliographystyle{IEEEtran}
\bibliography{references}

\end{document}